\pdfoutput=1
\documentclass[cernpreprint, UKenglish, texmf, orcidlogo]{atlasdoc}
\usepackage[subcaption, abbrevref=false]{atlaspackage}
\usepackage{atlasbiblatex}

\usepackage{atlasphysics}
\graphicspath{{logos/}{figs/}}

\usepackage{ANA-EXOT-2022-43-PAPER-defs}

\AtlasTitle{Search for single-production of vector-like quarks decaying into $Wb$ in the fully hadronic final state in $pp$ collisions at $\sqrt{s} = 13$~TeV with the ATLAS detector}

\AtlasAbstract{%
A search for $T$ and $Y$ vector-like quarks produced in proton--proton collisions at a centre-of-mass energy
of $13\,\text{TeV}$ and decaying into $Wb$ in the fully hadronic final state is presented.
The search uses $139~\text{fb}^{-1}$ of data collected by the ATLAS detector at the LHC from 2015 to 2018.
The final state is characterised by a hadronically decaying
$W$~boson with large Lorentz boost and a $b$-tagged jet, which are used to reconstruct the invariant mass of the vector-like quark candidate.
The main background is QCD multijet production, which is estimated using a data-driven method.
Upon finding no significant excess in data, mass limits at 95\% confidence level are obtained as a function of the global coupling parameter, $\kappa$.
The observed lower limits on the masses of  $Y$ quarks with $\kappa = 0.5$ and $\kappa = 0.7$ are
$2.0\,\text{TeV}$ and $2.4\,\text{TeV}$, respectively. For $T$ quarks,
the observed mass limits are $1.4\,\text{TeV}$ for $\kappa = 0.5$ and $1.9\,\text{TeV}$ for $\kappa = 0.7$.
}

\AtlasRefCode{EXOT-2022-43}

\PreprintIdNumber{CERN-EP-2024-222}

\arXivId{2409.20273}

\AtlasJournalRef{JHEP 02 (2025) 075}
\AtlasDOI{10.1007/JHEP02(2025)075}

\hypersetup{pdftitle={ATLAS document},pdfauthor={The ATLAS Collaboration}}

\begin{document}

\maketitle

\tableofcontents

\section{Introduction}
\label{sec:intro}

%

The discovery of the Higgs boson~\cite{HIGG-2012-27, CMS-HIG-12-028} at the Large Hadron Collider (LHC) and measurements of its width, in combination with previous experimental results, have significantly constrained the possibility of a
fourth generation of chiral quarks within the Standard Model of particle physics (SM)~\cite{Eberhardt:2012gv}. Although the SM has been tested to high precision, certain
theoretical questions still persist. One of them, the Higgs boson mass hierarchy problem, is related to the fact that the non-cancellable
radiative corrections to the Higgs boson mass are proportional to the square of the \enquote{cut-off} in the SM (the Planck scale, \qty{e19}{\GeV}),
whereas the observed Higgs boson mass is \qty{125}{\GeV}~\cite{tHooft:1980xb,Veltman}. The largest contributions to these corrections
originate from the loop interactions of the Higgs boson with the top quark. Various proposed beyond-the-SM (BSM)
theories solve this problem by cancelling out these corrections
via interactions with hypothetical particles. Vector-like quarks (VLQs) are one of those types of hypothetical spin-1/2 coloured multiplets in the $SU(3)_{C}\times SU(2)_{L} \times U(1)_{Y}$ representation. Unlike SM chiral quarks,
their left- and right-handed components transform similarly under SM gauge transformations. Consequently, since VLQs
do not acquire their mass via the Higgs mechanism, Higgs boson measurements do not constrain the
mass of these quarks. VLQs appear in BSM models such as the composite Higgs model~\cite{Kaplan:1983sm} and
little Higgs models~\cite{Arkani-Hamed:2001nha,Arkani-Hamed:2002ikv,Arkani-Hamed:2002sdy}. Phenomenological renormalisable VLQ models~\cite{Aguilar-Saavedra:2013qpa} predict the existence of
VLQ multiplets with fractional charges. In these models~\cite{Aguilar-Saavedra:2013qpa, DeSimone:2012fs, Matsedonskyi:2014mna, Buchkremer:2013bha}, VLQs couple primarily to the
third-generation SM quarks via the SM gauge bosons. Most VLQ searches in the current search
programme at the LHC probe the $qV$ decay modes, where
$V= W/Z/H \ \text{bosons}$ and $q = \text{top}\,(t)\ \text{or} \ \text{bottom}\,(b) \text{ quarks}$.

\begin{figure}[tb]
\centering
\includegraphics[scale=0.3]{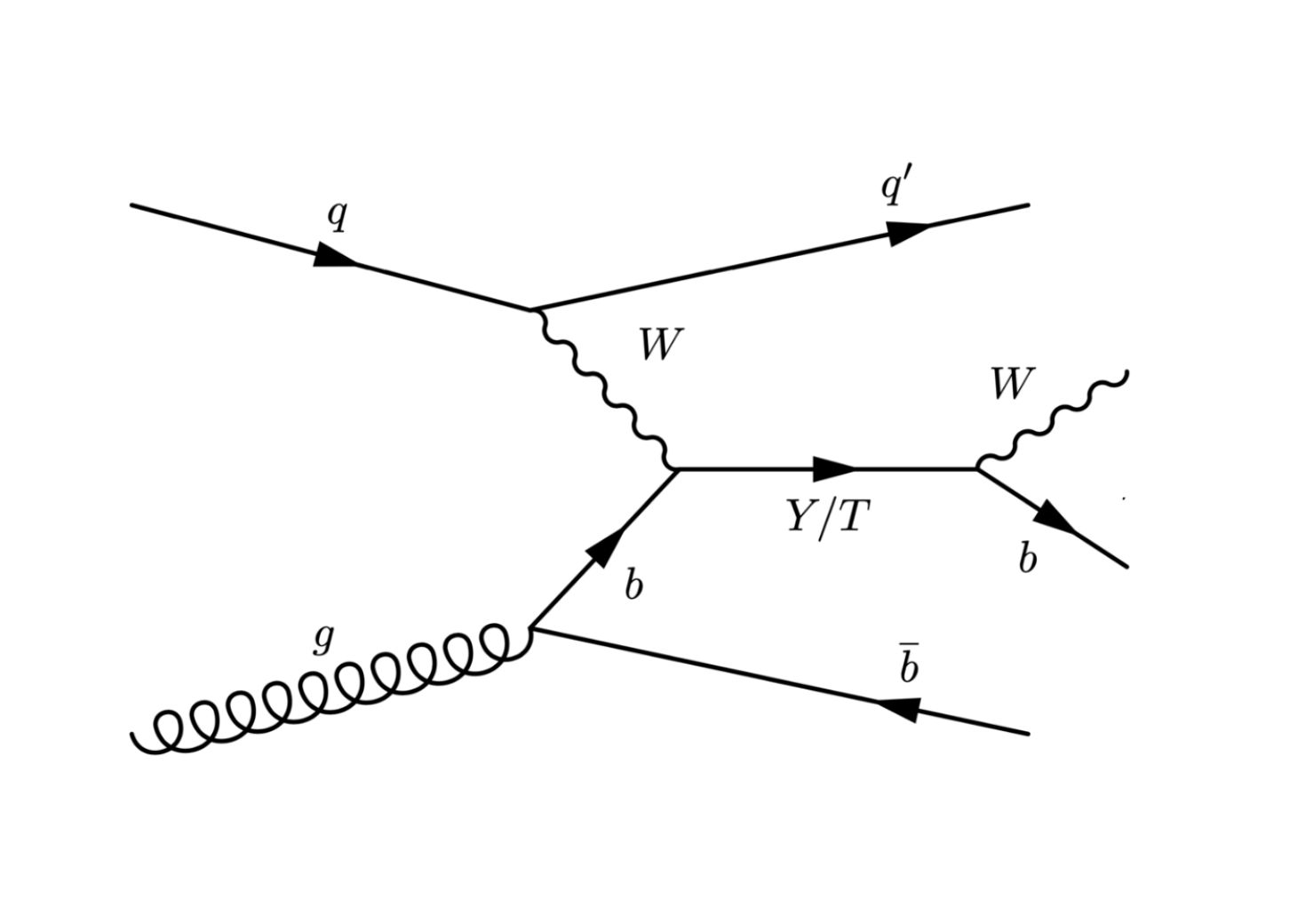}
\caption{Leading-order Feynman diagram for the $T/Y \rightarrow Wb$ process.}
\label{fig:feyn}
\end{figure}

In this paper, a search for $T$ and $Y$ VLQs decaying into a $Wb$ final state is performed using proton--proton ($pp$)
collisions in which the probed signal is produced via $Wb$ fusion. The $T$
VLQ has an electric charge of $+(2/3) e$ and occurs in singlet, doublet, and triplet representations, whereas the $Y$ VLQ with an electric charge of $-(4/3)e$ appears in doublet and triplet representations.
In the final state studied here, the $W$ boson decays hadronically.
Whereas previous searches for $pp\rightarrow T/Y(\rightarrow Wb)qb+X$
have used the semileptonic final state, this is the first search using the hadronic ($W \rightarrow q\bar{q}$)
decay mode. A typical lowest-order Feynman diagram for this signal process is shown in \cref{fig:feyn}.
In the mass range relevant for this paper, the contribution from the $t$-channel diagram is negligible for this process~\cite{Deandrea:2021vje}.

\FloatBarrier
The $Y$ quark decays solely
into a $Wb$ topology with a branching fraction ($\mathcal{B}$) of 100\%.
The $Y$ quark appears alongside the $B$ VLQ, which has an electric charge of $-(1/3)e$, in the $(B, Y)$ doublet and alongside both the $B$ and $T$ quarks in the $(T, B, Y)$ triplet. The $T$ singlet also
decays via the $Zt$ and $Ht$ modes. The relative branching fractions are mass dependent,
but for quark masses greater than \qty{1}{\TeV} they have a ratio of
$\mathcal{B} (Wb/Zt/Ht) = 2:1:1$~\cite{Aguilar-Saavedra:2013qpa}. The cross-section for pair-production processes
is higher at lower VLQ masses; however, single-production processes start to dominate at higher masses, e.g.\ around
\qty{1}{\TeV} for a coupling of 0.5. Additionally, since the production cross-section is proportional to the global coupling parameter $\kappa$, VLQ searches in the single-production mode can probe this coupling strength~\cite{Matsedonskyi:2014mna, Cacciapaglia:2011fx}.
The width ($\Gamma$) of a VLQ signal is also affected by $\kappa$.

The ATLAS and CMS collaborations have conducted extensive VLQ searches targeting both the single- and pair-production processes.
The most stringent lower limits of \qty{1.36}{\TeV} on the mass of pair-produced $T$ singlets arise from a $T\bar{T} \rightarrow Wb +X$ search performed using the \qty{13}{\TeV} $pp$ collision data collected in 2015--2018 by the
ATLAS Collaboration~\cite{EXOT-2019-06}. Similar single-production searches by ATLAS and CMS~\cite{EXOT-2016-17, EXOT-2019-07, EXOT-2018-52, EXOT-2020-01, EXOT-2022-40, EXOT-2021-02,  CMS-B2G-17-007,  CMS-B2G-18-003,  CMS-B2G-19-004,  CMS-B2G-21-007,  CMS-B2G-19-001} have set limits on both the mass of the VLQs
and model-dependent parameters, e.g.\ coupling strengths and mixing angles.
The limits set on the mixing parameter $\sin \theta_{\text{R}} $ ($\sin \theta_{\text{R}} = 2\cdot\sqrt{2} \cdot \kappa$) in the $(B,Y)$-doublet interpretation
in the $T/Y \rightarrow Wb$ search by ATLAS using 2015--2016 data are comparable to limits from
electroweak observables in the mass range between \qty{0.9}{\TeV} and \qty{1.25}{\TeV}~\cite{EXOT-2016-17}.
In the same search, the $Y$ quark was excluded for all masses less than \qty{1.7}{\TeV} for
$\kappa = 0.35$.
In a recent result from the ATLAS Collaboration with the 2015--2018 dataset, where a
statistical combination of
most $T \rightarrow Ht$ and $T \rightarrow Zt$ searches has been performed,
$\kappa > 0.5$ has been excluded for $m_{T}$ values up to \qty{2.2}{\TeV}~\cite{ EXOT-2021-02}.
The various limits quoted in this section are at 95\% confidence level (CL).

This search for $T/Y \rightarrow Wb$ is based on the full Run 2 dataset of $pp$ collisions at a centre-of-mass energy $\sqrt{s} = \qty{13}{\TeV}$
collected with the ATLAS detector during 2015--2018, corresponding to an integrated luminosity of
$\qty{139}{\per\femto\barn}$. The large integrated luminosity used allows probing a mass range for the $T$ and $Y$ VLQs from $\qty{1.0}{\TeV}$ to $\qty{2.7}{\TeV}$
and values of $\kappa$ from 0.1 to 1. Due to the high masses probed in this analysis, the $W$~boson from the VLQ decay is produced with large Lorentz boost.
The invariant mass of the VLQ candidate is reconstructed using the hadronically decaying boosted $W$~boson and $b$-tagged jet present in the final state.
Compared to the previous ATLAS search for $T/Y \rightarrow Wb$
in the single-lepton channel~\cite{EXOT-2016-17}, this search uses almost four times more integrated luminosity, and exploits a different decay mode, thus being complementary to it.

%
%
%
%
%
%
%
%
%
%
%
%
%
%
%
%


%
\section{ATLAS detector}
\label{sec:detector}

%
\newcommand{\AtlasCoordFootnote}{%
ATLAS uses a right-handed coordinate system with its origin at the nominal interaction point (IP)
in the centre of the detector and the \(z\)-axis along the beam pipe.
The \(x\)-axis points from the IP to the centre of the LHC ring,
and the \(y\)-axis points upwards.
Polar coordinates \((r,\phi)\) are used in the transverse plane,
\(\phi\) being the azimuthal angle around the \(z\)-axis.
The pseudorapidity is defined in terms of the polar angle \(\theta\) as \(\eta = -\ln \tan(\theta/2)\) and is equal to the rapidity
$ y = \frac{1}{2} \ln \left( \frac{E + p_z}{E - p_z} \right) $ in the relativistic limit.
Angular distance is measured in units of \(\Delta R \equiv \sqrt{(\Delta y)^{2} + (\Delta\phi)^{2}}\).}

The ATLAS detector~\cite{PERF-2007-01} at the LHC covers nearly the entire solid angle around the collision point.\footnote{\AtlasCoordFootnote}
It consists of an inner tracking detector surrounded by a thin superconducting solenoid, electromagnetic and hadronic calorimeters,
and a muon spectrometer incorporating three large superconducting air-core toroidal magnets.

The inner-detector system (ID) is immersed in a \qty{2}{\tesla} axial magnetic field
and provides charged-particle tracking in the range \(|\eta| < 2.5\).
The high-granularity silicon pixel detector covers the vertex region and typically provides four measurements per track,
the first hit generally being in the insertable B-layer (IBL) installed before Run~2~\cite{ATLAS-TDR-19,PIX-2018-001}.
It is followed by the SemiConductor Tracker (SCT), which usually provides eight measurements per track.
These silicon detectors are complemented by the transition radiation tracker (TRT),
which enables radially extended track reconstruction up to \(|\eta| = 2.0\).
The TRT also provides electron identification information
based on the fraction of hits (typically 30 in total) above a higher energy-deposit threshold corresponding to transition radiation.

The calorimeter system covers the pseudorapidity range \(|\eta| < 4.9\).
Within the region \(|\eta|< 3.2\), electromagnetic calorimetry is provided by barrel and
endcap high-granularity lead/liquid-argon (LAr) calorimeters,
with an additional thin LAr presampler covering \(|\eta| < 1.8\)
to correct for energy loss in material upstream of the calorimeters.
Hadronic calorimetry is provided by the steel/scintillator-tile calorimeter,
segmented into three barrel structures within \(|\eta| < 1.7\), and two copper/LAr hadronic endcap calorimeters.
The solid angle coverage is completed with forward copper/LAr and tungsten/LAr calorimeter modules
optimised for electromagnetic and hadronic energy measurements respectively.

The muon spectrometer (MS) comprises separate trigger and
high-precision tracking chambers measuring the deflection of muons in a magnetic field generated by the superconducting air-core toroidal magnets.
The field integral of the toroids ranges between \num{2.0} and \qty{6.0}{\tesla\metre}
across most of the detector.
Three layers of precision chambers, each consisting of layers of monitored drift tubes, cover the region \(|\eta| < 2.7\),
complemented by cathode-strip chambers in the forward region, where the background is highest.
The muon trigger system covers the range \(|\eta| < 2.4\) with resistive-plate chambers in the barrel, and thin-gap chambers in the endcap regions.

The luminosity is measured mainly by the LUCID--2~\cite{LUCID2} detector that records Cherenkov light produced in the quartz windows of photomultipliers located close to the beampipe.

Events are selected by the first-level trigger system implemented in custom hardware,
followed by selections made by algorithms implemented in software in the high-level trigger~\cite{TRIG-2016-01}.
The first-level trigger accepts events from the \qty{40}{\MHz} bunch crossings at a rate below \qty{100}{\kHz},
which the high-level trigger further reduces in order to record complete events to disk at about \qty{1}{\kHz}.

A software suite~\cite{SOFT-2022-02} is used in data simulation, in the reconstruction
and analysis of real and simulated data, in detector operations, and in the trigger and data acquisition
systems of the experiment.


%
\section{Event reconstruction}
\label{sec:reco}

%

This paper describes a search for the process $T/Y \rightarrow Wb$ where the final state is characterised by a hadronic decay of the boosted $W$ boson. The $T/Y$ VLQ is reconstructed using its decay products:
a large-radius (large-$R$) jet and a small-radius (small-$R$) jet, both with high transverse momentum (\pT).

Primary interaction vertices are reconstructed from at least two tracks with \pT > \qty{0.5}{\GeV}. For events with multiple primary vertex candidates, the one where the corresponding tracks have largest
sum of squared \pT is chosen~\cite{ATL-PHYS-PUB-2015-026}.

Large-$R$ jets are reconstructed from three-dimensional topological clusters~\cite{PERF-2014-07} of energy deposits in the calorimeter using the anti-$k_t$ algorithm~\cite{Cacciari:2008gp, Fastjet}
with radius parameter $R = 1.0$. These jets are calibrated to the hadronic energy scale with the local cluster weighting procedure~\cite{PERF-2014-07}.
The jets are then trimmed to reduce contributions from multiple interactions in the same and neighbouring bunch
crossings (\pileup) and soft interactions by reclustering the constituents of the jet
into $R = 0.2$ subjets with the $k_t$ algorithm~\cite{Catani:1993hr, Ellis:1993tq} and removing constituents
originating from subjets with \pT less than 5\% of the \pT of the parent jet. The four-momentum of the large-$R$ jet
is recomputed from the four-momenta of the remaining constituents and corrected using SM Monte Carlo simulation and
data~\cite{JETM-2018-02}. The events used in this search were
selected by a set of triggers requiring at least one such
large-$R$ jet~\cite{TRIG-2012-01}. This search uses
large-$R$ jets with \pT > \qty{500}{\GeV} and $|\eta| < 2.0$ and hence achieves maximum efficiency for these triggers.

Small-$R$ jets are also reconstructed with the anti-$k_t$ algorithm, but using $R =0.4$ and input constituents constructed by a particle-flow algorithm from calorimeter energy clusters and corresponding tracks in the ID~\cite{PERF-2015-09}. Jets are calibrated to the particle level using corrections derived from simulation and {\textit{in situ}} measurements~\cite{JETM-2018-05}.
In this analysis, both \enquote{central} ($|\eta| < 2.5$) and \enquote{forward} ($ 2.5 < |\eta| < 4.5$) small-$R$ jets are used. For the central small-$R$ jets with \pT < \qty{60}{\GeV}, a jet-vertex tagging algorithm is employed to mitigate \pileup by
ensuring the jets match the primary collision vertex~\cite{PERF-2014-03}. The jet-vertex tagging is also performed for the forward jets~\cite{ATL-PHYS-PUB-2019-026}.
The jet-vertex tagger's efficiency, as measured in simulated events,
differs from that measured in data and thus the efficiency ratio is used as a per-event correction factor.

The large-$R$ jet is identified as a $W$ boson using a three-variable cut-based $W$-tagger~\cite{JETM-2018-03} which is defined using jet substructure variables.
The tagger is optimised for two working points (WPs), which are chosen to achieve signal efficiencies of 50\% (tight) and 80\% (loose), respectively~\cite{JETM-2018-03}. Both WPs are used in this analysis.
Tagging scale factors which quantify the relative tagging-efficiency differences between data and simulation are derived
for both signal ($W$-tagged jets) and background jets ($\gamma+$jets\,/\,multijets)~\cite{ATL-PHYS-PUB-2020-017}.
The signal scale factors are evaluated using \ttbar events in the lepton-plus-jets final state.

In each event, central small-$R$ jets which contain $b$-hadrons are identified using the DL1r tagging algorithm~\cite{FTAG-2019-07}.
This tagger is based on a deep neural network that utilises impact-parameter information from tracks associated with the jet, as well as topological properties of their secondary and tertiary vertices.
A 70\% efficiency working point for identifying true $b$-jets in \ttbar events is utilised in this paper. Scale factors are applied to
account for remaining small tagging-efficiency discrepancies between data and simulation in this analysis~\cite{FTAG-2018-01, FTAG-2020-08, FTAG-2019-02}.

This analysis vetoes events containing charged leptons (electrons or muons). Electron candidates are reconstructed from energy deposits in the
EM calorimeter matched to charged-particle tracks in the inner detector~\cite{EGAM-2018-01}.
The electrons are required to have \pT > \qty{25}{\GeV} and $|\eta| < 2.47$, excluding the transition region between barrel and endcap calorimeters
($1.37 < |\eta| < 1.52$), and pass the \enquote{tight} identification criteria~\cite{EGAM-2018-01}. Muon candidates are
reconstructed from matching tracks in the ID and the MS~\cite{MUON-2018-03}.
The muons are required to have \pT > \qty{25}{\GeV} and $|\eta| < 2.5$, and satisfy the \enquote{medium} identification and \enquote{tight} isolation criteria~\cite{MUON-2018-03}.


%
\section{Simulated event samples}
\label{sec:modelling}

%

Monte Carlo (MC) simulation events are employed to model signal and background distributions.
They undergo full ATLAS detector simulation using \GEANT~\cite{Agostinelli:2002hh}
or a faster simulation with parameterised calorimeter showers~\cite{SOFT-2010-01}.
In-time and out-of-time \pileup effects were simulated by overlaying minimum-bias interactions generated with \PYTHIA[8.186]~\cite{Sjostrand:2007gs}, adjusted to match observed \pileup conditions. \EVTGEN~\cite{Lange:2001uf} modelled heavy-flavour hadron decays, except for processes generated with \SHERPA~\cite{Gleisberg:2008ta}.
Simulated events undergo the same reconstruction and analysis as data events, with small corrections to object selection efficiencies, energy scales, and resolutions are applied for better agreement with data.
The main simulation settings for the MC events are discussed in this section.
Signal events were generated using the faster simulation,
while all SM background MC simulated events were generated with a detailed \GEANT model of the ATLAS detector.

\subsection{Simulated signal events}

Signal events with single $T$-quark production were simulated at leading-order (LO) with the \MGNLO[2.3.3]
event generator~\cite{Alwall:2014hca} using the \NNPDF[2.3lo]~\cite{Ball:2012cx} parton distribution function (PDF) set.
The samples are normalised by multiplying the LO cross-section times branching fraction for given assumed couplings
by a correction factor to account for finite-width effects~\cite{Roy:2020fqf,Roy:2022wjj}, and by a $K$-factor to correct the LO cross-section to the next-to-leading-order (NLO) prediction computed in
the narrow-width approximation~\cite{Cacciapaglia:2018qep}.
The event generator was interfaced with \PYTHIA[8.212]~\cite{Sjostrand:2014zea} to model parton showering, hadronisation, and the underlying event with the \NNPDF[2.3lo]~\cite{Ball:2012cx} PDF set and the A14~\cite{ATL-PHYS-PUB-2014-021} set of tuned parameters (tune).
The matrix-element calculations are based on the phenomenological model described in Ref.~\cite{Buchkremer:2013bha}, which includes all tree-level processes. It is
assumed that the VLQs couple exclusively to SM quarks of the third generation. Events were
generated for $T (\rightarrow Wb) qb$ processes at fixed values of $\kappa$ and the $T$-quark mass from \qtyrange{1.0}{2.7}{\TeV}.
During the generation process, matrix-element-based event weights~\cite{Alwall:2011uj} were calculated. These weights are subsequently used to reweight the events in each sample to different mass ($m$) and $\kappa$ values, effectively creating a grid in the $m{-}\kappa$ plane.
The overall
acceptance times efficiency for $T$-quark signal events with a mass of \qty{1.6}{\TeV} and $\kappa = 0.5$ is 1.7\%,
following the kinematic cuts in the signal region described in~\cref{sec:event_sel}. The Y-quark signals were not simulated separately, as the Y-quark distributions can be obtained by multiplying
the $T$-quark yield by a factor of two, to account for the larger branching fraction to $Wb$.

\subsection{Simulated background events}
The main background in this search comes from QCD multijet production. It is modelled using a data-driven method, with MC-based corrections to account for possible correlations inherent in the method, as detailed in~\cref{sec:ABCD}.
The other SM background contributions are estimated using MC simulations.

The modelling of \ttbar events utilised the next-to-leading-order (NLO) \POWHEGBOX[v2]~\cite{Nason:2004rx,Frixione:2007vw,Alioli:2010xd}
generator and the \NNPDF[3.0nlo]~\cite{Ball:2014uwa} PDF set.
The events were then interfaced with \PYTHIA[8.230], which employed the
\NNPDF[2.3lo] PDF set and the A14 tune.
To regulate the effects of high-\pT radiation and achieve appropriate matrix element to parton shower matching, the \hdamp parameter in
\POWHEGBOX was set to $1.5\,m_t$~\cite{ATL-PHYS-PUB-2017-007}, where $m_t = \qty{172.5}{\GeV}$.
Corrections were applied to the top-quark kinematics to account for NLO electroweak effects and next-to-next-to leading-order (NNLO)
QCD effects~\cite{Czakon:2019bcq}.
To ensure proper normalisation, the events were scaled to the cross-section computed at NNLO in QCD,
incorporating the resummation of next-to-next-to-leading logarithmic soft gluon terms, using
\TOPpp[2.0]~\cite{Cacciari:2011hy,Beneke:2011mq,Baernreuther:2012ws,Czakon:2012zr,Czakon:2012pz,Czakon:2013goa,Czakon:2011xx}.

The generation of single-top-quark events involving $Wt$-, $t$-, and $s$-channel processes utilised \POWHEGBOX[v2] and the \NNPDF[3.0nlo] PDF set.
Parton showering, hadronisation, and the underlying event were modelled by \PYTHIA[8.230], using the \NNPDF[2.3lo] PDF set and the A14 tune.
To address interference between the \ttbar and $Wt$ final states, the \enquote{diagram removal} (DR) scheme~\cite{Frixione:2005vw, Frixione:2008yi}
was employed.  Uncertainties in modelling the interference are estimated by comparing this sample with an alternative sample generated with the \enquote{diagram subtraction} (DS) scheme~\cite{Frixione:2008yi,White:2009yt}.

The \SHERPA[2.2.8] generator~\cite{Bothmann:2019yzt} was used to model $W$/$Z$+jets events.
The matrix-element calculation incorporates up to two partons at NLO and up to four partons at LO.
The merging of the matrix-element calculation with the \SHERPA parton shower was achieved using the \MEPSatNLO prescription~\cite{Hoeche:2012yf}.
The \NNPDF[3.0nnlo]~\cite{Ball:2014uwa} PDF set was used for the matrix-element calculation.  To estimate the modelling uncertainty,
an alternative $W$+jets sample was generated using \SHERPA[2.2.8], where the
modelling of hadronisation is based on the Lund
string model~\cite{Andersson:1983ia, Sjostrand:1984ic}.


%
\section{Analysis strategy}
\label{sec:event_sel}

%

This analysis searches for $T/Y$ VLQs decaying into a  $W$ boson and a $b$-quark, where the $W$ boson decays hadronically.
Since the probed VLQs are in the \unit{\TeV} mass range, the $W$ boson is boosted and thus its decay products  have low angular separation and are reconstructed as a single large-$R$ jet.
The main background comes from QCD multijet production and is estimated using a data-driven method. For this estimation, four control regions, one validation region, and one signal region are defined.
The reconstructed invariant mass of the VLQ (\VLQM) is the discriminating variable for this analysis.
For each event in the six analysis regions, \VLQM is defined as the magnitude of the vector sum of the four momenta of the
leading (highest-\pT) large-$R$ jet and the leading small-$R$ jet that has an angular separation of $\Delta R > 1$ from the leading large-$R$ jet.
This is motivated by the fact that a $T/Y$ VLQ with mass greater than \qty{1}{\TeV} would decay into two objects with a large angular separation.
The validation region is used to validate the data-driven estimate of the multijet background.
Finally, the estimated multijet background, and other SM backgrounds estimated using MC simulation, are
used in a binned maximum-likelihood fit in the signal region to search for a $T/Y$ signal
in the $m{-}\kappa$ plane.

\subsection{Event preselection}

Events of interest are required to have no charged leptons, as defined in \cref{sec:reco},  and at least one large-$R$ jet and at least one small-$R$ jet that do not overlap
with each other.  The definition of \VLQM for each event ensures an angular separation of $\dR > 1$ for the leading small-$R$ jet and the leading large-$R$ jet.
Signal events are expected to be characterised by the presence of high-\pT jets in the final state. The leading large-$R$ jet is required to have $\pT > \qty{500}{\GeV}$
and the leading central small-$R$ jet is required to have $\pT > \qty{350}{\GeV}$. Events are required to contain at least one forward jet with $\pT > \qty{40}{\GeV}$. This ensures considerable suppression of the multijet background.
These kinematic
cuts are henceforth referred to as the \enquote{preselection}.

\subsection{Event categorisation}

Events passing the preselection are separated into six analysis regions: four control regions, one validation region, and one signal region.
The control and validation regions are defined for the sole purpose of estimating the pre-fit multijet background using the data-driven method as described in~\cref{sec:ABCD}.
The control and validation regions are not used in the statistical fits.

\begin{figure}[tb]
\centering
\includegraphics[scale=0.4]{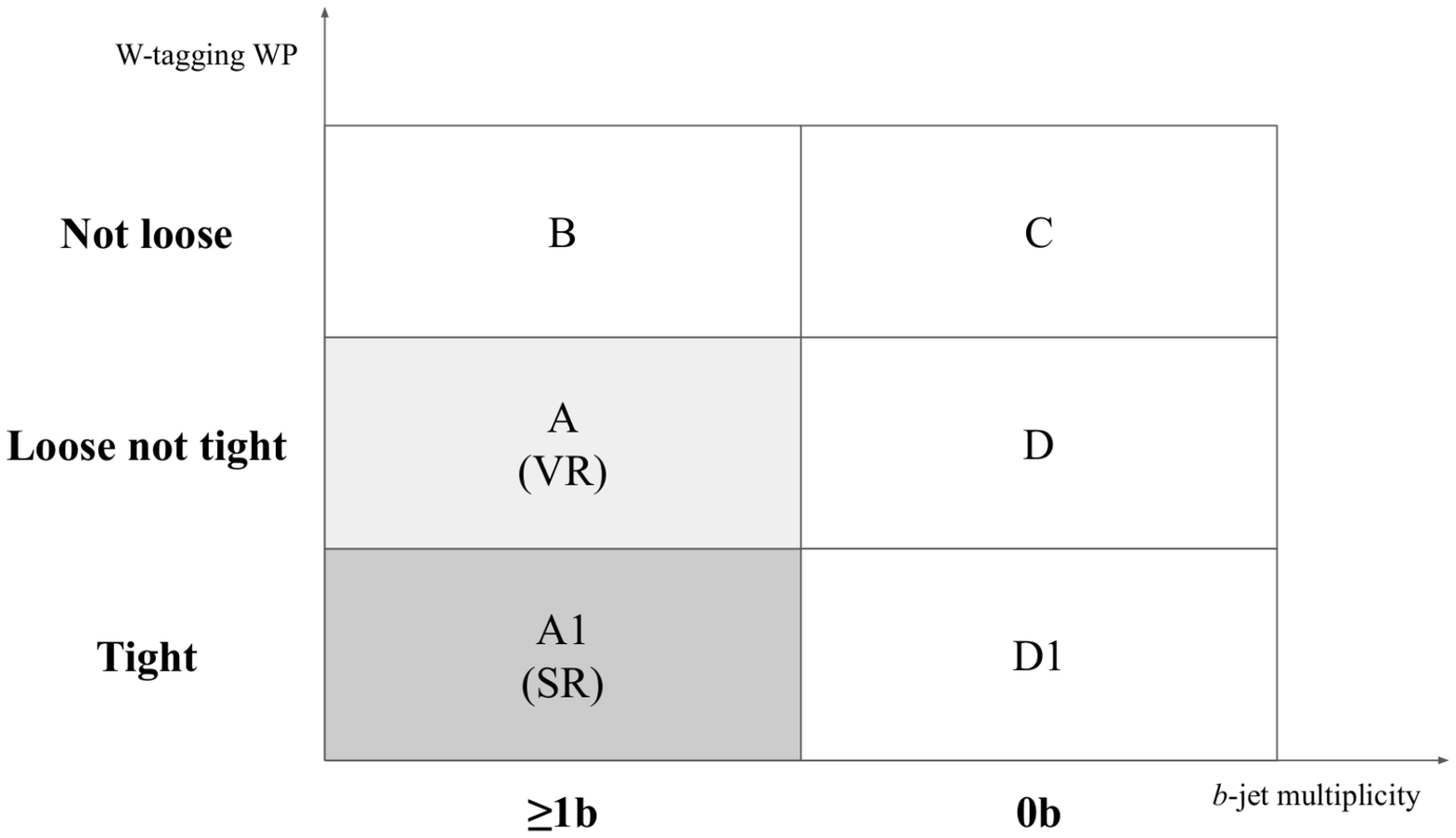}
\caption{A representation of the signal, validation, and control regions.
Region A1 is the signal region; region A is the validation region and the other four regions B, C, D, and D1 are the control regions.}
\label{fig:abcd:method:regions}
\end{figure}

The region categorisation is defined with two
orthogonal kinematic variables, namely the $b$-jet multiplicity and the $W$-tagging WPs of the three-variable $W$-tagger, as shown in~\cref{fig:abcd:method:regions}.
The $b$-jet multiplicity is calculated using $b$-jets defined in~\cref{sec:reco}.
Although the variables are expected to be nearly orthogonal, any departure from this ideal is accounted for in the analysis.

\FloatBarrier

\paragraph*{Signal region}
Events in the signal region (SR) A1 in~\cref{fig:abcd:method:regions} have at least one $b$-jet and the
leading large-$R$ jet is $W$-tagged using the tight WP (50\% tagging efficiency).
The strict requirement on the tagging efficiency improves the suppression of
light-flavour multijet background events.

\paragraph*{Validation region}
The validation region (VR) A is defined so as to be kinematically close to the signal region, yet
orthogonal to it. Events in region A similarly contain at least one $b$-jet but the leading large-$R$ jet is required to be loosely but not tightly $W$-tagged,
i.e.\ these events contain a leading large-$R$ jet that is $W$-tagged at the 80\% WP but not at the
50\% WP, thus making VR A orthogonal to SR A1.

\paragraph*{Control regions} The four control regions (CR) B, C, D, and D1 are used to constrain the shape and normalisation of the \VLQM distribution of the background processes
in the SR A1 and VR A. The CR B contains only
events which have at least one $b$-jet and a leading large-$R$ jet that is not $W$-tagged by the loose selection (80\% WP).
The three CRs C, D, and D1 contain only events that do not have any $b$-jets. These three
regions are distinguished by their $W$-tagging WP requirements.
The event yields in the four regions are ranked in descending order as follows: C, B, D, and D1.

For two different $T$-quark signals, one with mass \qty{1.1}{\TeV} and $\kappa = 0.25$ and the other with  mass \qty{1.5}{\TeV} and $\kappa = 0.45$,
the signal-to-background ratio is highest in the signal region and negligible in the validation and control regions.
The $S/B$ and $S/\sqrt{B}$ ratios for a \qty{1.1}{\TeV} signal with $\kappa = 0.25$ are 0.026 and 3.9 in the SR A1 and 0.005 and 1.8 in the VR A, respectively.
For a $T$-quark signal with mass \qty{1.5}{\TeV} and $\kappa = 0.45$, the $S/B$ and $S/\sqrt{B}$ ratios are 0.048 and 5.2 in the SR A1 and 0.008 and 2.5 in the VR A, respectively.
In all four CRs, the $S/B$ and $S/\sqrt{B}$ ratios for these two simulated signals are less than 0.1 and 1, respectively.

%
%
%
%
%
%
%
%
%
%
%
%
%
%
%
%


%
\section{Data-driven estimation of the multijet background}
\label{sec:ABCD}

%

QCD multijet production is the dominant source of background in this search. The shape and
normalisation of the \VLQM template for multijet events in the SR A1 and VR A in~\cref{fig:abcd:method:regions}
are estimated using a data-driven method similar to those used in Refs.~\cite{EXOT-2017-02,EXOT-2018-48,EXOT-2019-07}. In this method, two weakly correlated variables are used to construct the control regions.
The central assumption behind the success of this method is that these control regions are pure in multijet events and have negligible non-multijet SM background contamination.
Here, the multijet contribution in each bin of the \VLQM distribution
in SR A1 (VR A) is estimated from the multijet background in control regions
B, C, and D1 (D). The projected estimate of the multijet background in a given bin $i$
of the \VLQM distribution results from the subtraction of SM MC backgrounds (\ttbar, single-top-quark, $W/Z$+jets) from the observed data events in each corresponding bin.
Thus, the data-driven estimate for each bin $i$ in SR A1 (VR A) can be written as
\begin{equation*}
N_{\text{A/A1}}^{\text{multijet estimate}}[i] = \R\left[ i \right] \times (N_{\text{B}}^{\text{Data}}[i]-N_{\text{B}}^{\text{SM MC backgrounds}}[i]) \times \frac{(N_{\text{D/D1}}^{\text{Data}}[i]-N_{\text{D/D1}}^{\text{SM MC backgrounds}}[i])}{(N_{\text{C}}^{\text{Data}}[i]-N_{\text{C}}^{\text{SM MC backgrounds}}[i])}\,.
\label{eqn:abcd:correctionfactor}
\end{equation*}
The bin-by-bin correction factor $\R\left[ i \right]$ reflects the possible correlation between the two kinematic variables,
i.e.\ the $b$-jet multiplicity and $W$-tagging WP.
The $\R\left[ i \right]$ for SR A1 are calculated from the bin-by-bin ratios of simulated multijet events in the
\VLQM distributions for the regions specified below in Eq.~(\ref{eqn:Rdef}).
The overall normalisation of the \VLQM distribution of the
simulated multijet events in each of the four regions (A1, B, C, and D) is scaled to the estimated yield in data.
This scaling is done using two single-bin likelihood fits. In the first fit, the overall
normalisation of the simulated multijet events is scaled to the data yields by fixing
the normalisation of other SM backgrounds for the \VLQM distributions in regions A1 and B simultaneously.
Similarly, the second fit fixes the normalisation of those SM backgrounds in regions C and D1 simultaneously.
The normalisation scaling corrects for any mismodelling in the simulated multijet samples.
The correction factor for the signal region A1 can be written as
\begin{equation}
\R\left[ i \right]= \frac{\N{A1}{R}[i]}{\N{B}{R}[i]} \times \frac{\N{C}{R}[i]}{\N{D1}{R}[i]}\,.
\label{eqn:Rdef}
\end{equation}
Here, $\N{A1/B/C/D1}{R}$ denotes the normalisation-scaled multijet yields in the four regions
that are used to calculate \R.
The corresponding $\R\left[ i \right]$ for VR A is calculated by replacing
$\N{A1}{R}$ with $\N{A}{R}$,
and $\N{D1}{R}$ with $\N{D}{R}$, in Eq.~(\ref{eqn:Rdef}).
For this calculation of \R for VR A, the overall normalisations of the simulated multijet events are scaled to data yields
by fitting regions A and B together and regions C and D together.
These two single-bin fits are performed using the methodology described above for the corresponding fits for $\R\left[ i \right]$ in SR A1.
Thus in Eq.~(\ref{eqn:Rdef}) the multijet events in each bin $i$ of regions A1 and B are scaled by one normalisation factor, and another normalisation
factor is applied bin-by-bin in regions C and D1.

The value of $\R\left[ i \right]$ impacts
the final multijet \VLQM distribution for both SR A1 and VR A. To mitigate any statistical fluctuations
in the \R distribution as a function of \VLQM, a third-degree polynomial fit to the
\R distribution is performed. A third-degree polynomial fit is chosen because it gives the best agreement between the data and the estimated background.
The uncertainties arising from this fit are treated as systematic uncertainties. The uncertainty due to the choice of polynomial is smaller than the leading uncertainties assigned to the data-driven estimate and is neglected.
The values of \R used in the final calculation of the multijet event yields for the \VLQM distributions vary between 0.85 and 0.65 for SR A1 and between 0.9 and 0.7 for VR A.
The final estimated yields for
the multijet background in SR A1 and VR A are shown in~\cref{table:results:estimation}. The uncertainties include systematic and statistical contributions as detailed in~\cref{sec:syst}.

\newenvironment{DIFnomarkup}{}{}
\begin{DIFnomarkup}

\begin{table}[tb]
\centering
\sisetup{separate-uncertainty,table-format=6.0(6), round-mode=places, round-precision=0}
\caption{Summary of observed and predicted yields in SR A1 and VR A.
For the non-multijet SM backgrounds the MC predicted yields are tabulated,
while for the multijet background
the yields from the data-driven estimate are shown. These yields are used as inputs to the statistical fits described in~\cref{sec:result}.
The uncertainties include systematic and statistical contributions as detailed in~\cref{sec:syst}. The statistical
and systematic uncertainties in the non-multijet SM have been propagated and assigned to the data-driven multijet estimate.}
\label{table:results:estimation}
\begin{tabular}{l |
S[table-format=5.0]@{$\,\pm\,$}S[table-format=3.0]
S[table-format=5.0]@{$\,\pm\,$}S[table-format=5.0]
}
\toprule
& \multicolumn{2}{c}{SR A1} & \multicolumn{2}{c}{VR A}\\
\midrule
Single top   & 660 & 570 & 400 & 330 \\
\Wjets   & 770 & 140 & 750 & 130 \\
\ttbar   & 381 & 49 & 709& 81 \\
\Zjets   & 187 & 33 & 277 & 44 \\
Data-driven multijet estimate   & 11220 & 660 & 69000 & 14000 \\
\midrule
Total background  & 13220 & 860 & 71000 & 14000 \\
\midrule
Data   & \multicolumn{1}{S[table-format=5.0]@{\phantom{$\,\pm\,$}}}{12923} & &
\multicolumn{1}{S[table-format=5.0]@{\phantom{$\,\pm\,$}}}{62409} & \\
\bottomrule
\end{tabular}
\end{table}
\end{DIFnomarkup}

\FloatBarrier


%
\section{Systematic uncertainties}
\label{sec:syst}

%

The modelling of kinematic variables such as \VLQM in the signal and simulated background processes as described in~\cref{sec:modelling}
is affected by experimental uncertainties associated with the reconstruction and calibration of the underlying physics objects.
These uncertainties in the simulated backgrounds are propagated to the data-driven multijet background estimate.
Additionally, uncertainties are assigned to the multijet background to account for variations in the \R values and in the statistical uncertainty propagation methods. This includes the use of Gaussian uncertainty propagation and the statistical
uncertainties originating from the fits used in the estimation of the multijet background. Cross-section uncertainties are included for all the simulated backgrounds.

\subsection{Experimental and theoretical uncertainties}

The uncertainty in the integrated luminosity of the 2015--2018 ATLAS dataset is estimated to be 1.7\%~\cite{ATLAS-CONF-2019-021}.
The uncertainty is derived from the baseline luminosity measurements made by the LUCID-2 detector~\cite{LUCID2}. %
For small-$R$ and large-$R$ jets, the uncertainties associated with the energy scale and resolution
are evaluated using a combination of simulated events and \emph{in situ} methods applied to the collected data~\cite{JETM-2018-02}.
In addition, a mass scale uncertainty is evaluated for the large-$R$ jets with a forward-folding technique
that uses fits to the $W$-boson and top-quark mass peaks~\cite{ATLAS-CONF-2020-022} in simulation and data.
A jet mass resolution uncertainty of 20\% is applied to large-$R$ jets by smearing the energy of the jets.
This uncertainty estimate has negligible impact on the final result.

Uncertainties are assigned to correction factors for differences in $b$-tagging and $W$-tagging efficiencies between simulated events and data events.
Flavour-dependent tagging uncertainties are evaluated using differences in tagging response for $b$-jets, $c$-jets, and light-flavour jets between simulation and data~\cite{FTAG-2018-01, FTAG-2020-08, FTAG-2019-02}. Since the high-\pT region is probed in this search, an additional extrapolation uncertainty
is used to account for tagging inefficiencies in the range
$\pT > \qty{400}{\GeV}$~\cite{FTAG-2018-01}. Uncertainties in the correction factors for $W$-tagging arise from uncertainties related to
the jet energy scale, as well as modelling uncertainties and statistical uncertainties~\cite{JETM-2018-03}.
The correction factors are extrapolated into the high-\pT region ($\pT > \qty{600}{\GeV}$) and hence the
corresponding uncertainties are also propagated.

\subsection{Modelling uncertainties for simulated backgrounds}
Modelling uncertainties are assigned to single-top-quark and \Wjets events as they are the irreducible backgrounds secondary to the multijet background.
Uncertainties related to initial-state radiation (ISR), final-state radiation (FSR), and the parton shower were assessed for the single-top samples.
Due to significant interference between the $Wt$ and $\ttbar$ processes, two different schemas were employed to generate the $Wt$ MC event samples.
The nominal sample uses the diagram removal method, while the variation sample uses the diagram subtraction method. The difference between these two samples is used as the uncertainty.
The ISR/FSR uncertainty is determined similarly to the uncertainty estimation for \ttbar events.
For the \Wjets events, an alternative sample, where the modelling of hadronisation is based on the Lund string model, is used.
Theoretical cross-section uncertainties in the normalisation of the simulated SM backgrounds are included. For the single-top-quark
backgrounds an uncertainty of $\pm 2.5\%$ is included, for \ttbar an  uncertainty of $+2.4\%/-3.5\%$ is assigned, while for $W/Z$+jets a $\pm 6\%$ uncertainty is applied.

\subsection{Modelling uncertainties for data-driven background estimates}

\begin{figure}[htp]
\centering
\includegraphics[scale=0.4]{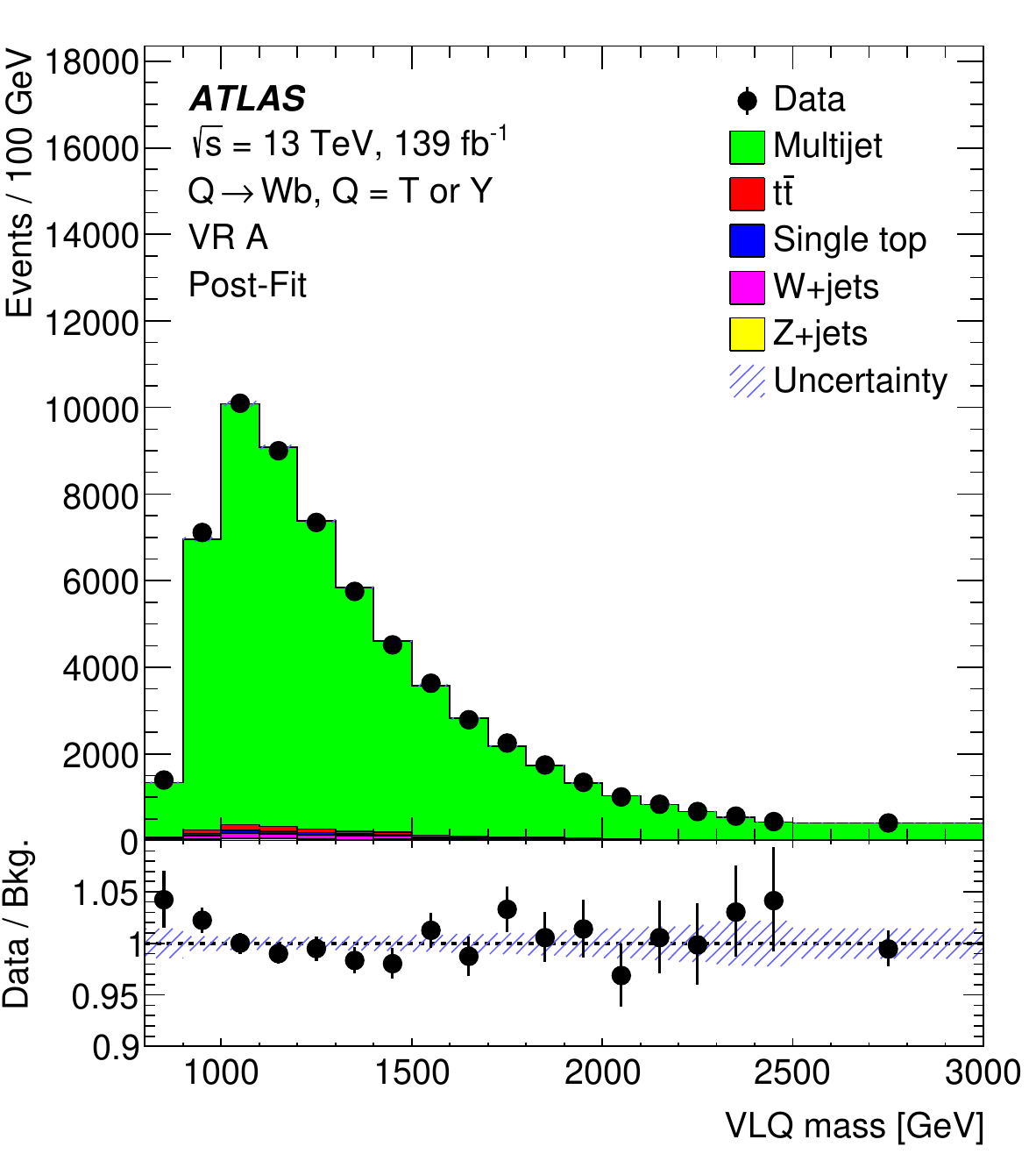}
\caption{The post-fit distribution for \VLQM in VR A after the fit to data under the background-only hypothesis.
All the uncertainties are included in this fit as nuisance parameters.
The lower panel depicts the ratio of data to the fitted background yields.
}
\label{fig:VR_postFit}
\end{figure}

\begin{figure}[htbp]
\centering
\includegraphics[scale=0.3]{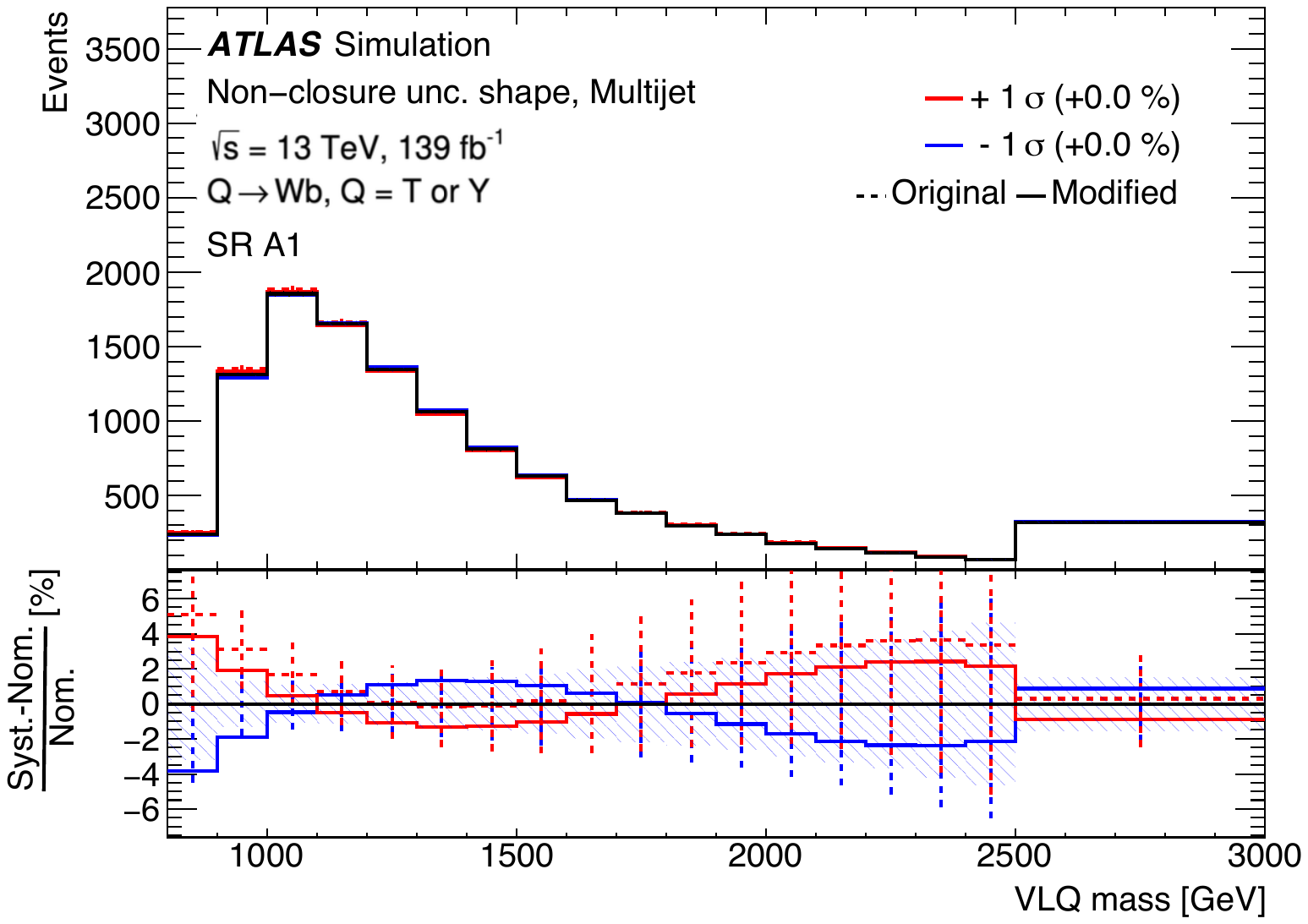}
\caption{SR shape of the non-closure uncertainty on the multijet background estimation. The (solid) modified uncertainty is the smoothed version of the uncertainty
that has been used in the likelihood fits and the original (dashed) uncertainty is the distribution before the smoothing procedure.}
\label{fig:MJ_VR_shape}
\end{figure}

The uncertainties in the data-driven method used to estimate the multijet background can be categorised according to their source.
First, experimental and theoretical uncertainties from the simulated SM backgrounds are propagated through the method, and variations in these backgrounds are used to estimate the uncertainties for the multijet estimate.
The variations of the simulated SM backgrounds are used in calculating the corresponding multijet event estimates for the \VLQM distributions.
Differences between the resulting \VLQM distributions of the multijet background
are used as corresponding uncertainties.
Statistical uncertainties arising from Gaussian propagation from the method's arithmetic operations are assigned as separate uncertainties.
The shape of the \R distributions is a function of the \Pythia[8] and \Sherpa dijet MC distributions used in the respective calculations.
Since the final data-driven estimate depends on these distributions,
uncertainties in the modelling of \R are evaluated by comparing the estimates of \R obtained by using dijet Monte Carlo samples from different generators, namely \Pythia[8] and \Sherpa.
The difference between the \Pythia[8] estimate and the \Sherpa estimate is used as an uncertainty. This uncertainty is larger in the VR than in the SR, resulting in a larger fractional uncertainty in the VR in~\cref{table:results:estimation}.
A third-degree polynomial is fitted to model \R for the multijet estimate, based on  the method described in~\cref{sec:ABCD}.
The statistical uncertainty arising from the polynomial fitting is also considered as a separate uncertainty.
An additional non-closure uncertainty is introduced to account for shape discrepancies
between the data and fitted backgrounds in the validation region.
The \VLQM distribution is fitted to data in the validation region under the background-only hypothesis.
The post-fit distribution from the fit is shown in~\cref{fig:VR_postFit}.
The shape of the non-closure between the data and the fitted backgrounds seen in the validation region in~\cref{fig:VR_postFit} is propagated as
a shape uncertainty for the multijet events in SR A1.
This uncertainty is introduced to account for a similar discrepancy that may arise in the fits to data in the SR. The shape of this uncertainty in SR A1 is shown in~\cref{fig:MJ_VR_shape}.
After injecting a nominal signal, it was checked that the distortion of the background shape is well below the shape uncertainties considered, so the expected bias is minimal.

\FloatBarrier
%

%
%
%
%
%
%
%
%
%
%
%
%
%
%
%


%
\section{Statistical analysis and results}
\label{sec:result}

%

A binned maximum-likelihood fit is performed on the \VLQM distribution in the SR A1 using the RooStats~\cite{Cowan:2010js} framework to test
the background-only hypothesis across the $m{-}\kappa$ plane for the VLQ
signals described in~\cref{sec:modelling}.  The background prediction used for all these fits is the aggregation of the data-driven multijet background
and the simulated SM backgrounds. The systematic uncertainties described in~\cref{sec:syst} are incorporated into the
likelihood as nuisance parameters $\theta_i$ via multiplicative Gaussian constraints, $G(\theta_{i})$~\cite{Conway:2011in}.
The normalisation of the multijet background, $\mu^{\text{multijets}}$,  is included as an unconstrained parameter of the fit.

\begin{figure}[tb]
\centering
\includegraphics[width=0.48\textwidth]{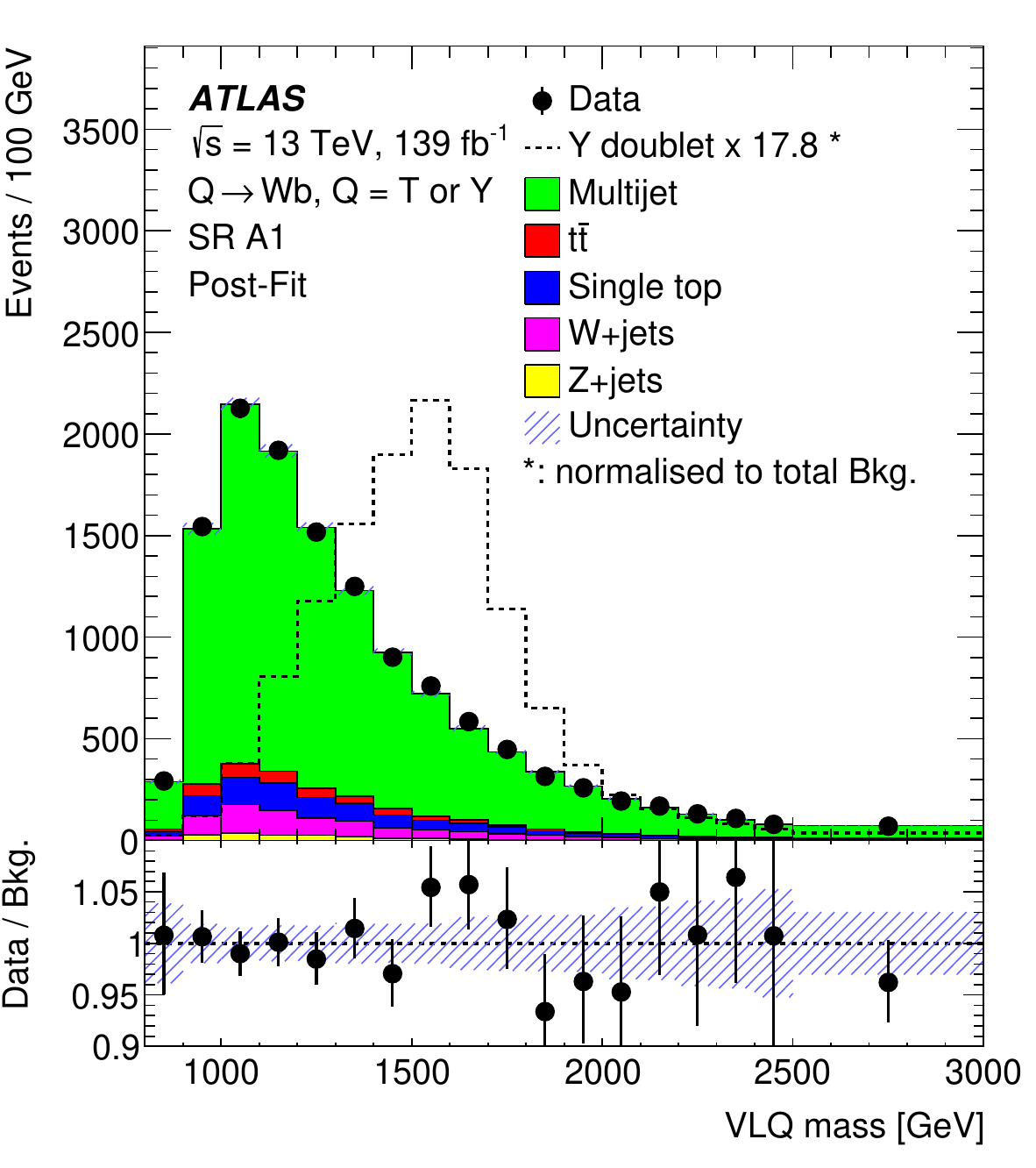}
\caption{The post-fit distribution of \VLQM in the SR A1 after the fit to data under the background-only hypothesis.
All the uncertainties are included in this fit as nuisance parameters.
The lower panel depicts the ratio of data to the fitted background yields.
The hatched area in the lower panel represents the total uncertainty in the background, including the uncertainty in $\mu^{\text{multijets}}$.
The overlaid dotted-line histogram in the upper panel shows the signal from a simulated $Y$ VLQ with mass \qty{1.6}{\TeV} and $\kappa=0.5$,
normalised to the total post-fit background yield.}
\label{fig:0lep_bonly}
\end{figure}

\begin{figure}[tb]
\centering
\subcaptionbox{\label{fig:k05}}%
{\includegraphics[width=0.48\textwidth]{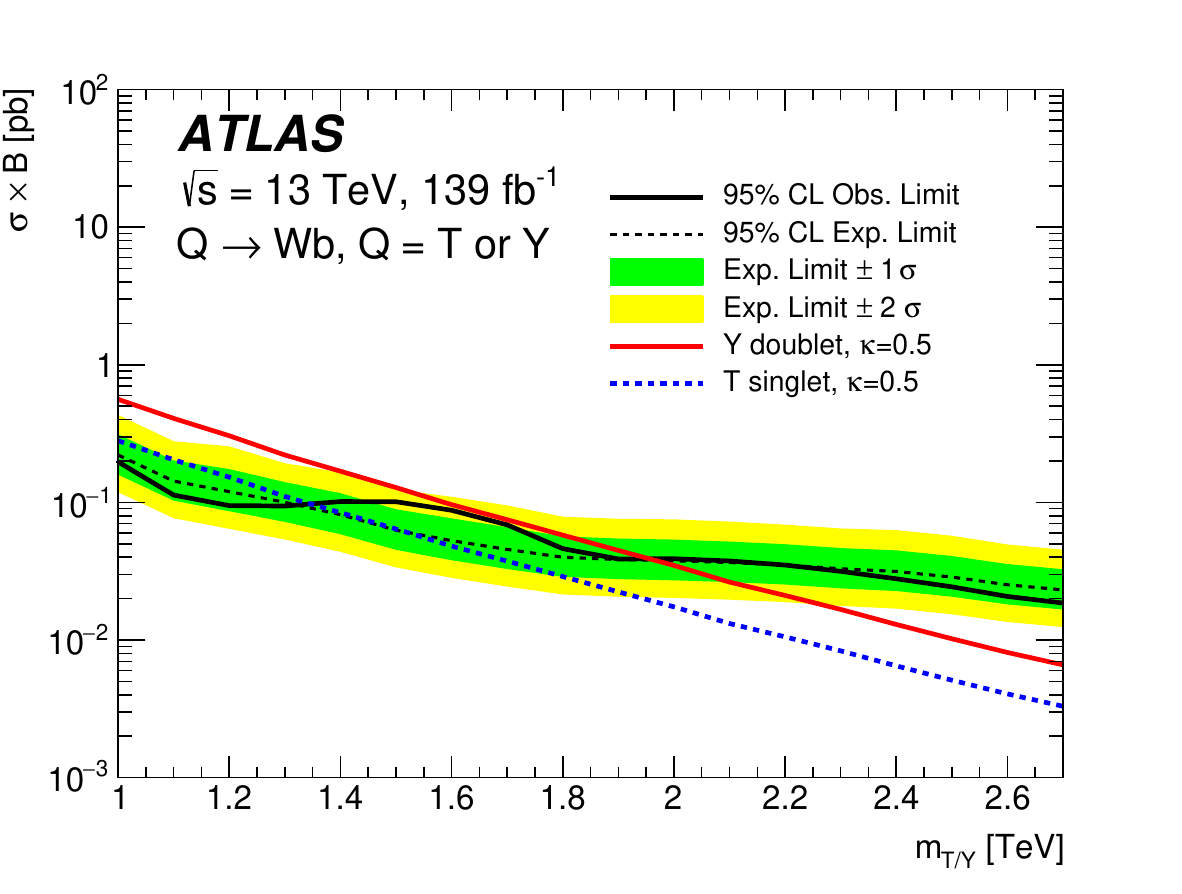}}%
\subcaptionbox{\label{fig:k07}}%
{\includegraphics[width=0.48\textwidth]{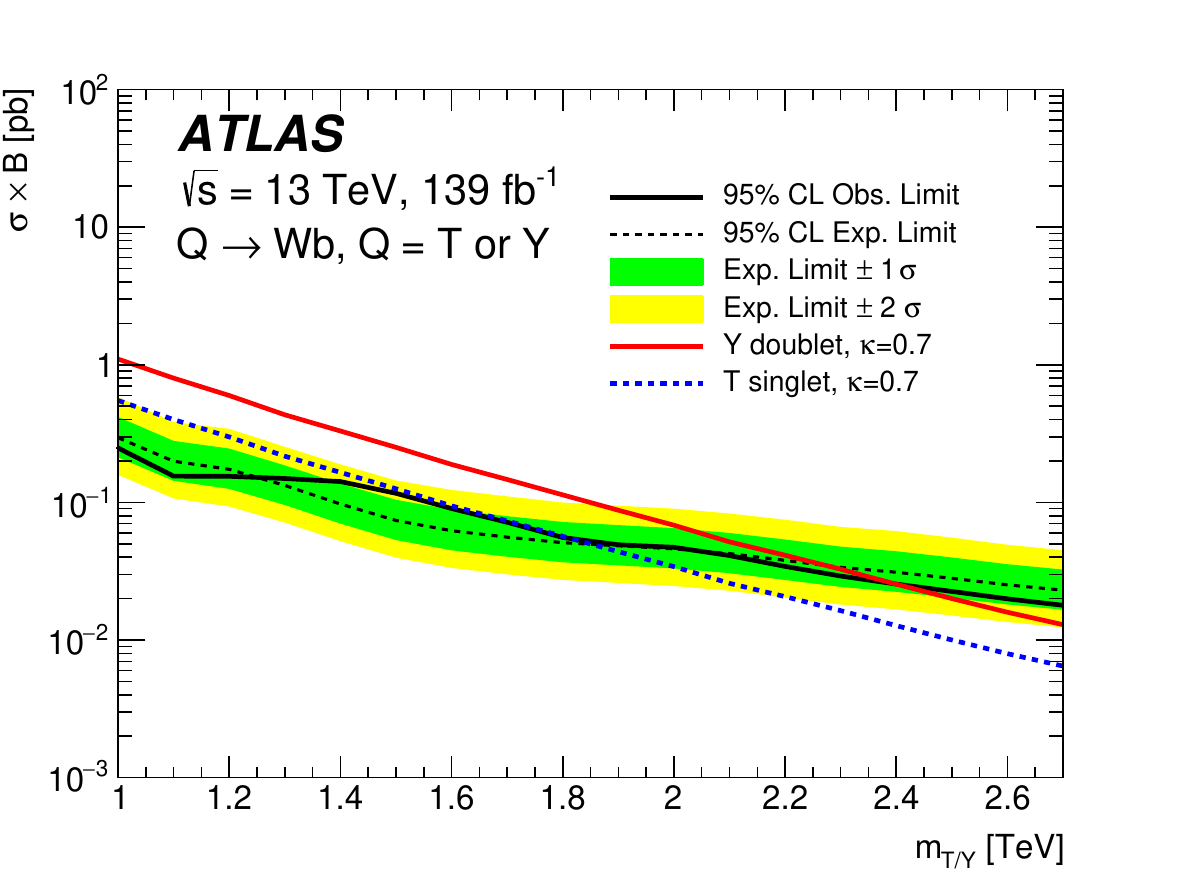}} \\
\caption{Expected (dotted) and observed (solid) cross-section limits times branching fraction for a $Y$ VLQ in the $(B,Y)$ doublet as a function of $m_{T/Y}$. Limits are computed for signals with couplings~\subref{fig:k05} $\kappa = 0.5$ and~\subref{fig:k07} $\kappa = 0.7$. The branching fraction $\mathcal{B}(Y \rightarrow Wb)$ is set to 1.
The surrounding bands correspond to $\pm 1$ and $\pm 2$ standard deviations around the expected limit. The $T$-singlet quark's theoretical cross-section (corrected to NLO with the inclusion of finite-width effects) times branching fraction ($\mathcal{B} (T \rightarrow Wb) = 0.5)$ as a function of the $T/Y$ mass is also shown (dashed blue line).
The mass limit for a $T$ singlet can then be obtained by computing the intersection of those
cross-section limits and the $T$-singlet theory cross-section curve.
The limits depend on $\kappa$ because the natural width of the $T/Y$ VLQ depends on $\kappa$.}
\label{fig:mass_limit_0lep}
\end{figure}

In the fit to data under the background-only hypothesis, $\mu^{\text{multijets}}$ is measured to be $\num{0.94 \pm 0.06}$.
The post-fit distribution of \VLQM in the SR A1 after the background-only fit to data is shown in~\cref{fig:0lep_bonly}.
The expected signal from a $Y$ VLQ with mass \qty{1.6}{\TeV} and $\kappa=0.5$, normalised to the total post-fit background yield, is overlaid.
The post-fit yields for the backgrounds and data are listed in~\cref{tab:postfit_0lep}.
The data agree with the background-only hypothesis and no significant excess of events above the SM predictions is observed.
A comparison of the uncertainties on the total background yields in~\cref{table:results:estimation} and~\cref{tab:postfit_0lep} imply
that the fit to data under the background-only hypothesis significantly constrains the leading pre-fit uncertainties on the total background yields.
In this fit, the relative post-fit impact of the systematic uncertainty on the inclusive background yield determined from the VLQ candidate invariant mass distribution for the non-closure uncertainty is 5.2\%,
and the impact of the uncertainty in the modelling of \R found by using \Sherpa as the alternative MC generator is 2.2\%.
A fit to data under the signal-plus-background hypothesis found a significance of $1.1 \sigma$ for a $T$-singlet signal with
mass \qty{1.6}{\TeV} and $\kappa = 0.5$.
In the absence of any significant excess, 95\% CL limits on the VLQ production cross-section are calculated separately for each signal hypothesis
in the $m{-}\kappa$ plane using the CL$_\text{s}$ method~\cite{Read:2002hq}.
The mass limits for $Y \rightarrow W(\rightarrow q\bar{q})b$ with $\kappa=0.5$ and $\kappa=0.7$ are shown in~\cref{fig:mass_limit_0lep}.
The limits depend on $\kappa$ because the total width of the \VLQM distribution varies with $\kappa$.
For $\kappa =0.5$, both the expected and observed mass limits for $Y$ quarks are \qty{2.0}{\TeV}, whereas for $\kappa = 0.7$ the expected and observed
limits are \qty{2.3}{\TeV} and \qty{2.4}{\TeV} respectively.
\Cref{fig:mass_limit_0lep} also shows the $T$ quark's theoretical cross-section times branching fraction ($\mathcal{B} (T \rightarrow Wb) = 0.5)$ as a function of its mass.
The limits for the $T$ singlet are weaker than those for $Y$ quarks in the $(B,Y)$-doublet representation.
The observed mass limits for $T$-singlet quarks as shown in~\cref{fig:mass_limit_0lep} are $\qty{1.4}{\TeV}$ for $\kappa=0.5$ and $\qty{1.9}{\TeV}$ for $\kappa=0.7$.
The limits in this search are affected more by systematic uncertainties than by statistical ones. The two uncertainties with the highest impact
on the exclusion limits affect the modelling of the data-driven multijet estimate, the first being the
non-closure uncertainty and the second being the uncertainty in the modelling of \R by using \Sherpa as the alternative MC generator.

\begin{DIFnomarkup}
\begin{table}[tb]
\caption{Post-fit background yields after the fit to data for the background-only hypothesis in the SR.
The quoted uncertainties in the yields include contributions from statistical and systematic sources, and are computed taking into account correlations among nuisance parameters resulting from the fit to data.}
\label{tab:postfit_0lep}
\centering
\begin{tabular}{l|S[table-format=5(3), table-alignment=right]}
\toprule
& {SR A1} \\
\midrule
Single top & 870 \pm 570  \\
$W$+jets & 780 \pm 140  \\
$Z$+jets & 190 \pm 32  \\
\ttbar & 388 \pm 48  \\
Multijet & 10690 \pm 530  \\
\midrule
Total background & 12920 \pm 180  \\
\midrule
Data & 12923  \\
\bottomrule
\end{tabular}
\end{table}
\end{DIFnomarkup}

\begin{figure}[htbp]
\centering
\includegraphics[scale=0.6]{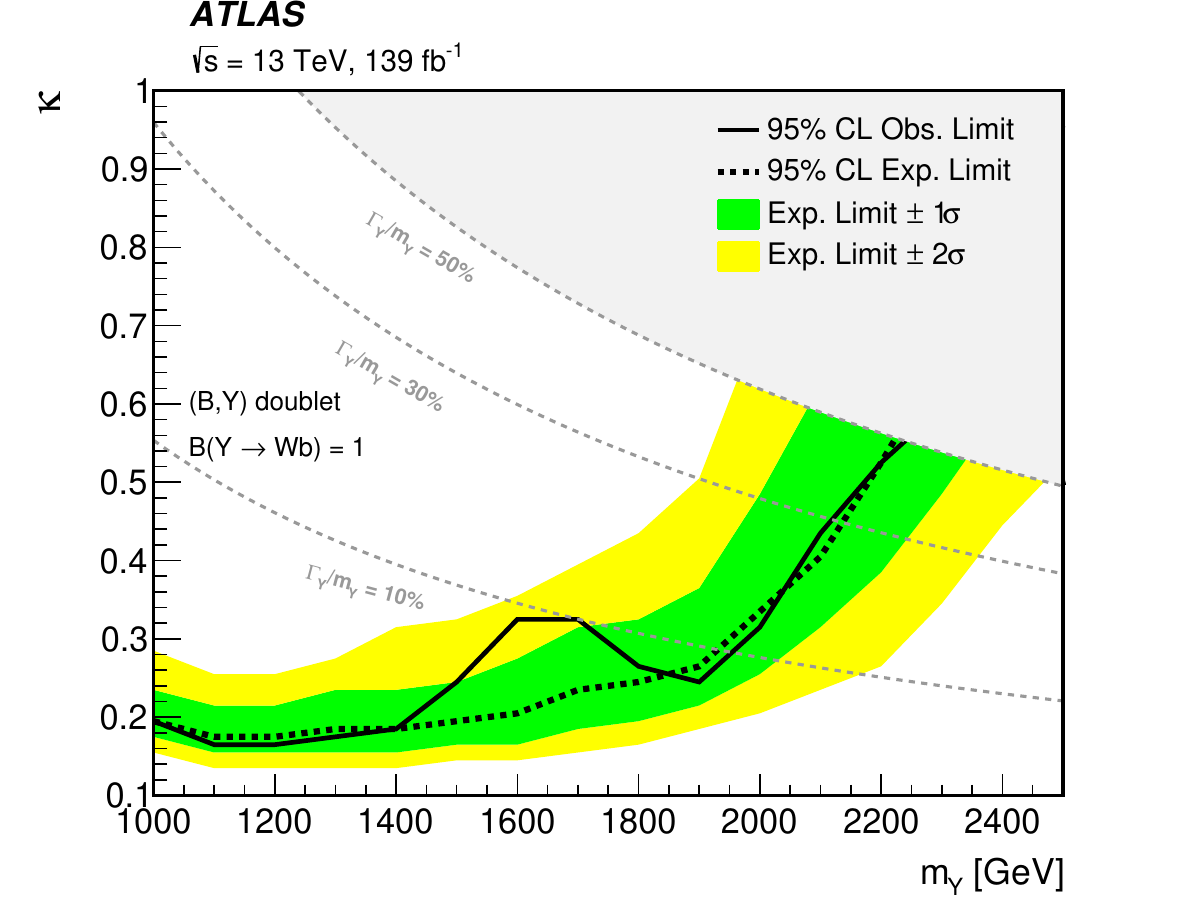}
\caption{Observed (solid) and expected (dotted) 95\% CL exclusion limits on the coupling constant $\kappa$ as a function of the $Y$ VLQ mass in the $(B,Y)$-doublet scenario.
All $\kappa$ values above the black contour lines are excluded at each mass point. The bands correspond to $\pm 1$ and $\pm 2$ standard deviations around the expected limit.
Limits are only presented in the regime $\Gamma_{Y}/m_{Y} \leq 50\%$, where
the theory calculations are known to be valid. The grey dotted isolines depict the highest $(m, \kappa)$ values allowed for
$Y$-doublet signals with various widths up to $\Gamma_{Y}/m_{Y} = 50\%$.}
\label{fig:2dplot}
\end{figure}

The results are also interpreted in a more generalised representation of the parameter space
in~\cref{fig:2dplot}, showing the largest excluded mass at a given $\kappa$.
Thus, for a coupling of $\kappa = 0.3$, $Y$-quark masses below \qty{1.5}{\TeV} are excluded.
For width-to-mass ratios less than 50\% in the $m_{Y}{-}\kappa$ plane, the $\Gamma_Y/m_Y$ isolines are displayed along with the two-dimensional limits.

\FloatBarrier


%

%
%
\FloatBarrier

\section{Conclusion}
\label{sec:conclusion}


A search for $T$ and $Y$ vector-like quarks decaying into a $Wb$ final state using proton--proton ($pp$) collisions
at a centre-of-mass energy of $\sqrt{s} = \qty{13}{\TeV}$, is presented in this paper.
The search uses $\qty{139}{\per\fb}$ of data recorded from 2015 to 2018 by the ATLAS experiment at the LHC.
Whereas previous ATLAS searches using this topology focused on the leptonic decay mode of the $W$ boson,
this is the first search where the hadronic decay mode of the $W$ boson is probed.
This analysis searches for VLQs with masses in the \unit{\TeV} range.
Consequently, the $W$ boson has a large Lorentz boost, and is reconstructed using advanced tagging algorithms
that use substructure information from large-$R$ jets.
The main background comes from QCD multijet production.
The shape and expected yield of this background are estimated using a data-driven method.
Upon finding no significant excess of events in data, mass limits as a function of the global coupling parameter $\kappa$ are determined.
The observed lower limits on the masses of $Y$ quarks with $\kappa = 0.5$ and $\kappa = 0.7$ are $\qty{2.0}{TeV}$ and $\qty{2.4}{TeV}$, respectively.
For $T$ quarks, the observed mass limits are $\qty{1.4}{TeV}$ for $\kappa = 0.5$ and $\qty{1.9}{TeV}$ for $\kappa = 0.7$.
All coupling values $\kappa > 0.6$ for $Y$-quark masses up to \qty{2.2}{\TeV}
in the $(B,Y)$ doublet are excluded for narrow-width signals ($\Gamma_{Y}/m_{Y} < 50\%$).
The mass limits on the $Y$ quark in the $(B,Y)$-doublet representation improve on
those of the previous ATLAS search, obtained using leptonic decays of the $W$ boson, by \qty{0.6}{\TeV} for $\kappa=0.5$.
Compared to the previous ATLAS search, this paper extends the search region for $Y$ quarks from \qty{2.0}{\TeV} to \qty{2.7}{\TeV}.
For the $T$-singlet case, with $T\rightarrow Wb$ decays, the search region is extended from \qty{1.2}{\TeV} to \qty{2.7}{\TeV}.
Depending on the choice of $\kappa$, the reported mass limits are the most stringent to date for this channel.


%

%
\section*{Acknowledgements}
%

%
%

%
%

We thank CERN for the very successful operation of the LHC and its injectors, as well as the support staff at
CERN and at our institutions worldwide without whom ATLAS could not be operated efficiently.

The crucial computing support from all WLCG partners is acknowledged gratefully, in particular from CERN, the ATLAS Tier-1 facilities at TRIUMF/SFU (Canada), NDGF (Denmark, Norway, Sweden), CC-IN2P3 (France), KIT/GridKA (Germany), INFN-CNAF (Italy), NL-T1 (Netherlands), PIC (Spain), RAL (UK) and BNL (USA), the Tier-2 facilities worldwide and large non-WLCG resource providers. Major contributors of computing resources are listed in Ref.~\cite{ATL-SOFT-PUB-2023-001}.

We gratefully acknowledge the support of ANPCyT, Argentina; YerPhI, Armenia; ARC, Australia; BMWFW and FWF, Austria; ANAS, Azerbaijan; CNPq and FAPESP, Brazil; NSERC, NRC and CFI, Canada; CERN; ANID, Chile; CAS, MOST and NSFC, China; Minciencias, Colombia; MEYS CR, Czech Republic; DNRF and DNSRC, Denmark; IN2P3-CNRS and CEA-DRF/IRFU, France; SRNSFG, Georgia; BMBF, HGF and MPG, Germany; GSRI, Greece; RGC and Hong Kong SAR, China; ISF and Benoziyo Center, Israel; INFN, Italy; MEXT and JSPS, Japan; CNRST, Morocco; NWO, Netherlands; RCN, Norway; MNiSW, Poland; FCT, Portugal; MNE/IFA, Romania; MSTDI, Serbia; MSSR, Slovakia; ARIS and MVZI, Slovenia; DSI/NRF, South Africa; MICIU/AEI, Spain; SRC and Wallenberg Foundation, Sweden; SERI, SNSF and Cantons of Bern and Geneva, Switzerland; NSTC, Taipei; TENMAK, T\"urkiye; STFC/UKRI, United Kingdom; DOE and NSF, United States of America.

Individual groups and members have received support from BCKDF, CANARIE, CRC and DRAC, Canada; CERN-CZ, FORTE and PRIMUS, Czech Republic; COST, ERC, ERDF, Horizon 2020, ICSC-NextGenerationEU and Marie Sk{\l}odowska-Curie Actions, European Union; Investissements d'Avenir Labex, Investissements d'Avenir Idex and ANR, France; DFG and AvH Foundation, Germany; Herakleitos, Thales and Aristeia programmes co-financed by EU-ESF and the Greek NSRF, Greece; BSF-NSF and MINERVA, Israel; NCN and NAWA, Poland; La Caixa Banking Foundation, CERCA Programme Generalitat de Catalunya and PROMETEO and GenT Programmes Generalitat Valenciana, Spain; G\"{o}ran Gustafssons Stiftelse, Sweden; The Royal Society and Leverhulme Trust, United Kingdom.

In addition, individual members wish to acknowledge support from Armenia: Yerevan Physics Institute (FAPERJ); CERN: European Organization for Nuclear Research (CERN PJAS); Chile: Agencia Nacional de Investigaci\'on y Desarrollo (FONDECYT 1230812, FONDECYT 1230987, FONDECYT 1240864); China: Chinese Ministry of Science and Technology (MOST-2023YFA1605700, MOST-2023YFA1609300), National Natural Science Foundation of China (NSFC - 12175119, NSFC 12275265, NSFC-12075060); Czech Republic: Czech Science Foundation (GACR - 24-11373S), Ministry of Education Youth and Sports (FORTE CZ.02.01.01/00/22\_008/0004632), PRIMUS Research Programme (PRIMUS/21/SCI/017); EU: H2020 European Research Council (ERC - 101002463); European Union: European Research Council (ERC - 948254, ERC 101089007), Horizon 2020 Framework Programme (MUCCA - CHIST-ERA-19-XAI-00), European Union, Future Artificial Intelligence Research (FAIR-NextGenerationEU PE00000013), Italian Center for High Performance Computing, Big Data and Quantum Computing (ICSC, NextGenerationEU); France: Agence Nationale de la Recherche (ANR-20-CE31-0013, ANR-21-CE31-0013, ANR-21-CE31-0022, ANR-22-EDIR-0002), Investissements d'Avenir Labex (ANR-11-LABX-0012); Germany: Baden-Württemberg Stiftung (BW Stiftung-Postdoc Eliteprogramme), Deutsche Forschungsgemeinschaft (DFG - 469666862, DFG - CR 312/5-2); Italy: Istituto Nazionale di Fisica Nucleare (ICSC, NextGenerationEU), Ministero dell'Università e della Ricerca (PRIN - 20223N7F8K - PNRR M4.C2.1.1); Japan: Japan Society for the Promotion of Science (JSPS KAKENHI JP22H01227, JSPS KAKENHI JP22H04944, JSPS KAKENHI JP22KK0227, JSPS KAKENHI JP23KK0245); Netherlands: Netherlands Organisation for Scientific Research (NWO Veni 2020 - VI.Veni.202.179); Norway: Research Council of Norway (RCN-314472); Poland: Ministry of Science and Higher Education (IDUB AGH, POB8, D4 no 9722), Polish National Agency for Academic Exchange (PPN/PPO/2020/1/00002/U/00001), Polish National Science Centre (NCN 2021/42/E/ST2/00350, NCN OPUS nr 2022/47/B/ST2/03059, NCN UMO-2019/34/E/ST2/00393, NCN \& H2020 MSCA 945339, UMO-2020/37/B/ST2/01043, UMO-2021/40/C/ST2/00187, UMO-2022/47/O/ST2/00148, UMO-2023/49/B/ST2/04085); Slovenia: Slovenian Research Agency (ARIS grant J1-3010); Spain: Generalitat Valenciana (Artemisa, FEDER, IDIFEDER/2018/048), Ministry of Science and Innovation (MCIN \& NextGenEU PCI2022-135018-2, MICIN \& FEDER PID2021-125273NB, RYC2019-028510-I, RYC2020-030254-I, RYC2021-031273-I, RYC2022-038164-I), PROMETEO and GenT Programmes Generalitat Valenciana (CIDEGENT/2019/027); Sweden: Carl Trygger Foundation (Carl Trygger Foundation CTS 22:2312), Swedish Research Council (Swedish Research Council 2023-04654, VR 2018-00482, VR 2022-03845, VR 2022-04683, VR 2023-03403, VR grant 2021-03651), Knut and Alice Wallenberg Foundation (KAW 2018.0157, KAW 2018.0458, KAW 2019.0447, KAW 2022.0358); Switzerland: Swiss National Science Foundation (SNSF - PCEFP2\_194658); United Kingdom: Leverhulme Trust (Leverhulme Trust RPG-2020-004), Royal Society (NIF-R1-231091); United States of America: U.S. Department of Energy (ECA DE-AC02-76SF00515), Neubauer Family Foundation.

%
%


%
%
%
%
%
%

%
%

%
%
%
\printbibliography
\clearpage
 
\begin{flushleft}
\hypersetup{urlcolor=black}
{\Large The ATLAS Collaboration}

\bigskip

\AtlasOrcid[0000-0002-6665-4934]{G.~Aad}$^\textrm{\scriptsize 104}$,
\AtlasOrcid[0000-0001-7616-1554]{E.~Aakvaag}$^\textrm{\scriptsize 17}$,
\AtlasOrcid[0000-0002-5888-2734]{B.~Abbott}$^\textrm{\scriptsize 123}$,
\AtlasOrcid[0000-0002-0287-5869]{S.~Abdelhameed}$^\textrm{\scriptsize 119a}$,
\AtlasOrcid[0000-0002-1002-1652]{K.~Abeling}$^\textrm{\scriptsize 56}$,
\AtlasOrcid[0000-0001-5763-2760]{N.J.~Abicht}$^\textrm{\scriptsize 50}$,
\AtlasOrcid[0000-0002-8496-9294]{S.H.~Abidi}$^\textrm{\scriptsize 30}$,
\AtlasOrcid[0009-0003-6578-220X]{M.~Aboelela}$^\textrm{\scriptsize 45}$,
\AtlasOrcid[0000-0002-9987-2292]{A.~Aboulhorma}$^\textrm{\scriptsize 36e}$,
\AtlasOrcid[0000-0001-5329-6640]{H.~Abramowicz}$^\textrm{\scriptsize 155}$
% ...
\bigskip
\\

% ...
$^{17}$Department for Physics and Technology, University of Bergen, Bergen; Norway.\\
% ...
$^{30}$Physics Department, Brookhaven National Laboratory, Upton NY; United States of America.\\
% ...
$^{36}$$^{(a)}$Facult\'e des Sciences Ain Chock, Universit\'e Hassan II de Casablanca;$^{(b)}$Facult\'{e} des Sciences, Universit\'{e} Ibn-Tofail, K\'{e}nitra;$^{(c)}$Facult\'e des Sciences Semlalia, Universit\'e Cadi Ayyad, LPHEA-Marrakech;$^{(d)}$LPMR, Facult\'e des Sciences, Universit\'e Mohamed Premier, Oujda;$^{(e)}$Facult\'e des sciences, Universit\'e Mohammed V, Rabat;$^{(f)}$Institute of Applied Physics, Mohammed VI Polytechnic University, Ben Guerir; Morocco.\\
% ...
$^{45}$Physics Department, Southern Methodist University, Dallas TX; United States of America.\\
% ...
$^{50}$Fakult\"{a}t Physik , Technische Universit{\"a}t Dortmund, Dortmund; Germany.\\
% ...
$^{56}$II. Physikalisches Institut, Georg-August-Universit\"{a}t G\"ottingen, G\"ottingen; Germany.\\
% ...
$^{104}$CPPM, Aix-Marseille Universit\'e, CNRS/IN2P3, Marseille; France.\\
% ...
$^{119}$$^{(a)}$New York University Abu Dhabi, Abu Dhabi;$^{(b)}$United Arab Emirates University, Al Ain; United Arab Emirates.\\
% ...
$^{123}$Homer L. Dodge Department of Physics and Astronomy, University of Oklahoma, Norman OK; United States of America.\\
% ...
$^{155}$Raymond and Beverly Sackler School of Physics and Astronomy, Tel Aviv University, Tel Aviv; Israel.\\
% ...
$^{a}$ Also Affiliated with an institute covered by a cooperation agreement with CERN.\\
$^{b}$ Also at An-Najah National University, Nablus; Palestine.\\
$^{c}$ Also at Borough of Manhattan Community College, City University of New York, New York NY; United States of America.\\
$^{d}$ Also at Center for High Energy Physics, Peking University; China.\\
$^{e}$ Also at Center for Interdisciplinary Research and Innovation (CIRI-AUTH), Thessaloniki; Greece.\\
$^{f}$ Also at CERN, Geneva; Switzerland.\\
% ...

\end{flushleft}


%

%
%

%

%
%
%
%

\end{document}